\documentstyle[epsf]{article}

\newcommand{\beq}{ \begin{equation} }
\newcommand{\eeq}{ \end{equation} }
\newcommand{\bea}{ \begin{eqnarray} }
\newcommand{\eea}{ \end{eqnarray} }
\newcommand{\be}{ \beta }
\newcommand{\f}{ \frac }
\newcommand{\de}{ \partial }
\newcommand{\vS}{ \vec{S} }
\newcommand{\vpi}{ \vec{\pi} }
\newcommand{\ex}{ {\rm e} }

\begin{document}
\thispagestyle{empty}
\parskip=12pt
\raggedbottom

\def\mytoday#1{{ } \ifcase\month \or
 January\or February\or March\or April\or May\or June\or
 July\or August\or September\or October\or November\or December\fi
 \space \number\year}
\noindent
\hspace*{9cm} BUTP--95/16\\
\hspace*{9cm} IFUP--TH 33/95\\
\vspace*{1cm}
\begin{center}
{\LARGE The absence of  cut--off effects for the fixed 
point action in 1--loop perturbation 
theory}\footnote{Work supported in part by Schweizerischer Nationalfonds.}

\vspace{1cm}

Federico Farchioni \\
Dipartimento di Fisica dell'Universit\`a and I.N.F.N.\\
Piazza Torricelli 2, I--56126 Pisa, Italy. 

\vspace{.5cm}

Peter Hasenfratz, 
Ferenc Niedermayer\footnote{On leave from the Institute of Theoretical
Physics, E\"otv\"os University, Budapest} and Alessandro 
Papa\footnote{On leave from Dipartimento di Fisica, Universit\`a di Pisa
and I.N.F.N., Pisa} \\
Institute for Theoretical Physics \\
University of Bern \\
Sidlerstrasse 5, CH--3012 Bern, Switzerland

\vspace{0.5cm}

\mytoday \\ \vspace*{0.5cm}

\nopagebreak[4]

\begin{abstract}
In order to support the formal renormalization group arguments that
the fixed point action of an asymptotically free model gives cut--off 
independent physical predictions in 1--loop perturbation theory, we
calculate the finite volume mass--gap $m(L)$ in the non--linear 
$\sigma$--model. No cut--off effect of the type $g^4\left(a/L\right)^n$
is seen for any $n$. The results are compared with those of the standard
and tree level improved Symanzik actions. 
\end{abstract}

\end{center}
\eject
 
\section{\bf Introduction}
\label{sec:intro}

Lattice actions lying on the renormalized trajectory of a renormalization 
group (RG) transformation are perfect in the sense that all the spectral
quantities are free of lattice artefacts independently of the resolution.
In asymptotically free theories the renormalized trajectory starts from
a fixed point (FP) at $g=0$, where $g$ is the asymptotically free 
coupling of the continuum formulation.
The FP action is the classically perfect lattice regularization of the
field theory \cite{HN}: its classical solutions (instantons) are
scale invariant and in quadratic approximation in the fields the spectrum 
is exact. The FP action performs amazingly well when used in numerical 
simulations at small correlation lengths also \cite{HN-II}.

Wilson remarked some time ago \cite{WIL} that --- according to formal RG
arguments --- the change of the FP action under a RG transformation
in 1--loop perturbation theory is simple:
\bea
\be\; S^{FP}\;\;\;\;\;\;\;\;\;\;\;\;
& -\!\!\! -\!\!\! -\!\!\! -\!\!\! - \!\!\! \longrightarrow &
\;\;\;\;\;\;\;\;\;\;\;\;\be^{\prime}\; S^{FP}\;\;, \label{eq:RG} \\
&\mbox{RG, 1--loop}& \nonumber
\eea
where $\be^{\prime}=\be-\Delta\be$ and $\Delta\be$ is fixed by the
first coefficient of the $\be$--function. Wilson did not elaborate
this problem further. In \cite{HN-I} a set of formal RG arguments were
presented to support the statement in eq.~(\ref{eq:RG}). The question
is important since eq.~(\ref{eq:RG}) would imply that the FP action
is 1--loop (quantum) perfect.

We are not able to make the formal arguments of \cite{HN-I} more 
rigorous\footnote{The statement in  eq.~(\ref{eq:RG}) is, presumably, not 
even strictly correct
due to possible redundant operators. They, however, would not change
the physical content of  eq.~(\ref{eq:RG}).}. We present here an explicit
1--loop calculation in the $d=2$ non--linear $\sigma$--model to
support the arguments in \cite{HN-I} further.

We calculate the mass gap $m(L)$ in 1--loop perturbation theory using 
the FP action constructed and studied in \cite{HN}. The mass gap $m(L)$ 
has already been  calculated up to 2 loops using the standard 
action \cite{LWW}. The 1--loop result in $O(N)$ has 
the general form\footnote{We denote the coupling constant in the action 
by $g^2$ deviating from the notation in \cite{HN}.}
\beq
m(L)\;=\;\f{N-1}{2}\;\f{1}{L}\;\left[\:g^2\:+\:g^4\left(\f{N-2}{2\pi}\;\ln
\f{L}{a} + R_1 + (N-1) R_2 \right)\;\right]\;\;,
\label{eq:MG}
\eeq
where $R_i$, $i = 1, 2$ are independent of $N$
\beq
R_i \; = \; A_i \; + \; \f{a^2}{L^2}\,\left( c_{i1} + d_{i1}\,\ln\f{L}{a}
\right)  + \; \f{a^4}{L^4} \, \left( c_{i2} + 
d_{i2}\,\ln\f{L}{a}  \right)   + \; \cdots\;\;\;\;.
\label{eq:Ri}
\eeq
The cut--off dependent terms are not universal, they depend on the 
explicit form of the lattice action. The constants $A_i$ determine
the relation between the coupling constants (or the $\Lambda$--parameters) 
of the different lattice regularizations.

If the relation (\ref{eq:RG}) is valid then no cut--off effects 
in $R_i$ should be present to arbitrary order in $(a/L)$.
We have calculated the terms $R_i$ in eq.~(\ref{eq:MG}) for the FP,
standard and tree level improved Symanzik actions \cite{SYM}. 
For the FP action
the coefficients $c_{ik}$ and $d_{ik}$ of the cut--off dependent terms 
in eq.~(\ref{eq:Ri})
turned out to be zero within the numerical precision of the calculation.
These coefficients are typically
${\cal O}(1)$ for the standard and Symanzik actions. 

Readers who are not interested in the technical details are advised to skip 
the next section and go directly to the results.

\section{\bf One--loop perturbation theory with the FP action}
\label{sec:1-loop}

The FP action is a specific lattice regularization of the formal expression
\beq
\be\;{\cal A}^{cont}(\vS)\;=
\;\f{\be}{2}\;\int d^2x\;\de_{\mu}\vS\:\de_{\mu}\vS\;\;,
\eeq
where $\be = 1/g^2$ and the $N$--component vector $\vS$ satisfies the 
constraint
$\vS^2(x) = 1$. It is convenient to parametrize the FP action as
\bea
{\cal A}^{FP}(\vS)\; &=&\;-\:\f{1}{2}\;\sum_{n,r}\;\rho(r)\;
\left(1-\vS_n\vS_{n+r}\right)\;+  \label{eq:par} \\
& & \sum_{n_1,n_2,n_3,n_4} c(n_1,n_2,n_3,n_4)\left(1-\vS_{n_1}\vS_{n_2}
\right)
\left(1-\vS_{n_3}\vS_{n_4}\right)\;+\;\cdots\;\;, \nonumber
\eea
where the coupling constants $\rho$, $c$, $\cdots\;\;$ are determined by 
a classical saddle point equation \cite{HN}. Writing
\beq
\vS_n\;=\;\left( \begin{array}{c}
\sqrt{1 - g^2\:\vpi^2_n}\\ g\:\vpi_n
\end{array} \right)\;\;,
\label{eq:spin}
\eeq
where the field $\vpi_n$ has $N-1$ components, a perturbation theory
can be set up by considering $g \vpi$ as a small fluctuation.
The higher order couplings, which are indicated only implicitly in  
eq.~(\ref{eq:par}), do not enter in a 1--loop calculation.

There is a  technical problem (which is independent of the action)
when the mass gap is calculated in a finite periodic box.
In order to obtain the mass gap the zero (spatial) momentum two--point 
function
is calculated at large time separations in a cylinder whose extension in 
time is much larger than $L$.
In this finite euclidean space there are $N-1$ zero modes which 
are, however easy to separate and handle \cite{PH}. The real problem is the 
presence of quasi--zero modes related to the slow motion of the 
``magnetization"
$\vec{\cal M}(t) = \sum_x \vS(t,x)$ \cite{BZ,HL,HNAF}. The dynamics  
of $\vec{\cal M}(t)$ is described in leading order by a rotator which
has the spectrum
\beq
E_l\;=\;\f{g^2}{2L}\;l(l + N - 2)\;\;,\;\;\;\;\;\; l=0, 1, \ldots\;\;\;\;\;
\;.
\label{eq:spectrum} 
\eeq
These are slow modes with energy much below the normal excitation energies
$\sim\:2\pi/L$. A possible solution is to introduce collective coordinates
for these quasi--zero modes and study their dynamics \cite{HNAF}. A more 
elegant and technically simpler
solution is to observe that free boundary conditions in time direction
project on $O(N)$ singlet states and so these modes enter only as 
intermediate states \cite{LWW}.

We consider a cylinder of size $(2T+1)\times L$ with free boundary conditions
at $x_0 = \pm\: T$, periodic boundary conditions in $x_1$, $\;x_1 = 0, 1, 
\cdots , L-1\;\;$\footnote{The lattice unit $a$ is put to $1$.} and 
calculate the
correlation function
\beq
C(\tau)\;=\;\f{1}{L^2}\;\sum_{x_1,y_1}\;\langle \; \vS(x)\:\vS(y) \; 
\rangle_{x_0 = -y_0 = \tau}\;\;\;.
\label{eq:corr}
\eeq
We shall stay close to the notations introduced in \cite{LWW}. The form of
the FP action eq.~(\ref{eq:par}) and the need to use free boundary conditions
suggest to work in configuration space.

The propagator has the form\footnote{Equations (9--13) are diagonal in the 
internal indices
$i, j = 2, \cdots, N$ and they are not given explicitly.}
\beq
D(x_0,x_0^{\prime}\:;\:x_1-x_1^{\prime})\;=\;\f{1}{L}\;
\sum_q\;\ex^{iq(x_1-x_1^{\prime})} \; R^{-1}(x_0,x_0^{\prime}\:;\:q)\;\;,
\label{eq:prop}
\eeq
where $q=2\pi /L \cdot k$, $k=0,\ldots,L-1$ 
and the $(2T+1)\times (2T+1)$ dimensional matrix 
$R(x_0,x_0^{\prime}\:;\:q)$ ($q$ fixed) is defined as
\beq
R(x_0,x_0^{\prime}\:;\:q)\;=\;\rho(x_0-x_0^{\prime}\:;\:q) \; - \; 
\delta_{x_0,x_0^{\prime}}f(x_0)\;\;\;,
\;\;\;\;\; q\neq0\;\;,
\label{eq:R}
\eeq
with 
\beq
f(x_0)\;=\;\sum_{x_0^{\prime}=-T}^{T}\;\rho(x_0-x_0^{\prime}\:;\:q=0)
\;\;\;.
\label{eq:f}
\eeq
In Eqs.~(\ref{eq:R}), (\ref{eq:f}) $\rho(x_0\:;\:q)$ are 
the quadratic couplings 
Fourier transformed in space
\beq
\rho(x_0\:;\:q)\;=\;\sum_{x_1=0}^{L-1}\;\ex^{{-iq x_1}}\;\rho(x_0,x_1)\;\;.
\label{eq:rho}
\eeq
For $q=0$ $R$ has an extra term (related to the constraint $\sum_{x}\vpi(x) = 0$ which enters
the path integral when eliminating the global zero mode)
\beq
R(x_0,x_0^{\prime}\:;\:q=0)\;=\;\rho(x_0-x_0^{\prime}\:;\:q=0) \; - \;
\delta_{x_0,x_0^{\prime}}f(x_0) \; + \;  \lambda \;\;,
\label{eq:exterm}
\eeq
where $\lambda$ is an arbitrary positive parameter. The limit 
$\lambda\rightarrow\infty$
corresponds to the constraint $\delta\left(\sum_{x}\;\vpi(x)\right)$, 
but it is easy to see
that the final results are independent of $\lambda$. Due to the free 
boundary conditions the propagator is not translation invariant in time.

Using the explicit representation of the quadratic couplings $\rho$ in 
eq.~(17)
of Ref.~\cite{HN} (with the optimized parameter $\kappa=2$) one can obtain 
the  propagator and those vertices which are proportional to $\rho$ to high 
precision (close to
machine precision). On the other hand, in solving the FP equations for 
$c(n_1,n_2,n_3,n_4)$ we had to introduce cuts.
The numerical errors in our results are dominated by the errors in the 
couplings $c$.

\section{\bf Results}
\label{sec:results}

In order to simplify the discussion and save space we present the results
for $N=3$. We introduce the notations $A = A_1 + 2 A_2$, $c_1 = c_{11} + 
2 c_{21}$, etc.

The constant $A$ can be calculated simply by using the 1--loop results on 
$m(L)$ in continuum perturbation theory in the $\overline{\mbox{MS}}$ scheme
\cite{LUE}, and the ratios between $\Lambda_{\overline{MS}}$ and
the $\Lambda$--parameter of the lattice action under consideration.
For the standard and the Symanzik actions this ratio is known \cite{PAR}
\cite{IWA}. Using the general expression in the Appendix of \cite{HN}
we obtained for the FP action 
\beq
\Lambda_{FP}^{(N=3)} \; = \; 9.424754598 \; \Lambda_{st} \;\;\; ,
\label{eq:lambda}
\eeq
where $\Lambda_{FP}$ is the $\Lambda$--parameter of the action defined
by the couplings $\rho$ and $c$ of eq.~(\ref{eq:par}) which are used 
in the following mass gap calculation. The number in eq.~(\ref{eq:lambda})
is somewhat different from that corresponding to a parametrized form of the 
FP action which was used in numerical simulations earlier \cite{HN}.
The corresponding constants $A$ are given in Table~1 for the standard,
Symanzik and FP actions.

\begin{table}[htb]

\begin{center}
TABLE 1 \\
The constant $A = A_1 + (N-1)A_2 $ (see eq.~(\ref{eq:Ri})) \\
is given for $N=3$ for the different actions considered.
\end{center}

\centering

\begin{tabular}{cccc} \cline{1 - 4}
 & \hspace{.4in} standard  \hspace{.4in} & 
   \hspace{.4in} Symanzik  \hspace{.4in} & 
   \hspace{.4in}    FP     \hspace{.4in} \\
  \cline{1 - 4} 
 $A$  &  0.214836206  & 0.087964307   & $-0.142202395$  \\ 
\cline{1 - 4}
\end{tabular}

\end{table}

We calculated the two point function $C(\tau)$ of eq.~(\ref{eq:corr})
on a cylinder $(2T+1) \times L$, where $\tau\cdot 4\pi/L \gg 1$ and
$(T - \tau)\cdot 4\pi/L \gg 1$, where $4\pi/L$ is the energy
of the first excited state in the singlet channel \cite{LWW}.
Depending on $L$ we used $T$ and $\tau$ in the range $40 - 90$ and $9 - 45$,
respectively. The consistency conditions assuring correct exponentialization
\cite{LWW} were satisfied up to 9 digits, or better. The cut--off dependent
part of the ${\cal O}(g^4)$ result $(R - A)$ is given in Table~2 
for the different actions and $L=2, 3, \ldots, 10$ .

\begin{table}[htb]

\begin{center}
TABLE 2 \\
$(R - A)$ for the standard, the tree level Symanzik improved \\ 
and the FP actions for various L.
\end{center}

\centering

\begin{tabular}{cccc} \cline{1 - 4}
 $L$  
& \hspace{.25in} $R_{st}-A_{st}$  \hspace{.25in} 
& \hspace{.25in} $R_{Sym}-A_{Sym}$\hspace{.25in} 
& \hspace{.25in} $R_{FP}-A_{FP}$  \hspace{.25in} \\ \cline{1 - 4}
 2 & 0.086457646  & 0.020052439 & $-0.000302928$  \\
 3 & 0.037574920  & 0.004628052 & $-0.000011974$  \\
 4 & 0.020334020  & 0.001463304 & $-0.000003526$  \\
 5 & 0.012697662  & 0.000607401 & $-0.000001045$  \\
 6 & 0.008699585  & 0.000298780 & $-0.000000314$  \\
 7 & 0.006343465  & 0.000163770 & $-0.000000103$  \\
 8 & 0.004834753  & 0.000097027 & $-0.000000036$  \\
 9 & 0.003808864  & 0.000061034 & $-0.000000008$  \\
10 & 0.003078960  & 0.000040271 & $-0.000000005$  \\
\cline{1 - 4}
\end{tabular}

\end{table}

Both the standard and the Symanzik actions give power decaying cut--off 
corrections. In the latter case the $\sim a^2/L^2$ leading term seems
to be missing, or very small. For the FP action the power--like cut--off
effects are tiny, in the range of $L = 5, \ldots , 10$ are about 5 
orders of magnitude smaller than those of the standard action. 
The numerical errors in the results of the FP action are dominated 
by the errors in the quartic couplings $c$. They are obtained by
solving the FP equation by iteration where unavoidably cuts have to be 
introduced \cite{HN}. For the quartic couplings one can derive different
sum rules which are satisfied by our couplings up to 6--digits
accuracy. We did not attempt to translate this error into a quantitative
error estimate on the mass gap $m(L)$, but it seems to us plausible that it
can produce the tiny power--like cut--off effects seen.

\begin{figure}[htb]
\begin{center}
\leavevmode
\epsfxsize=90mm
\epsfbox{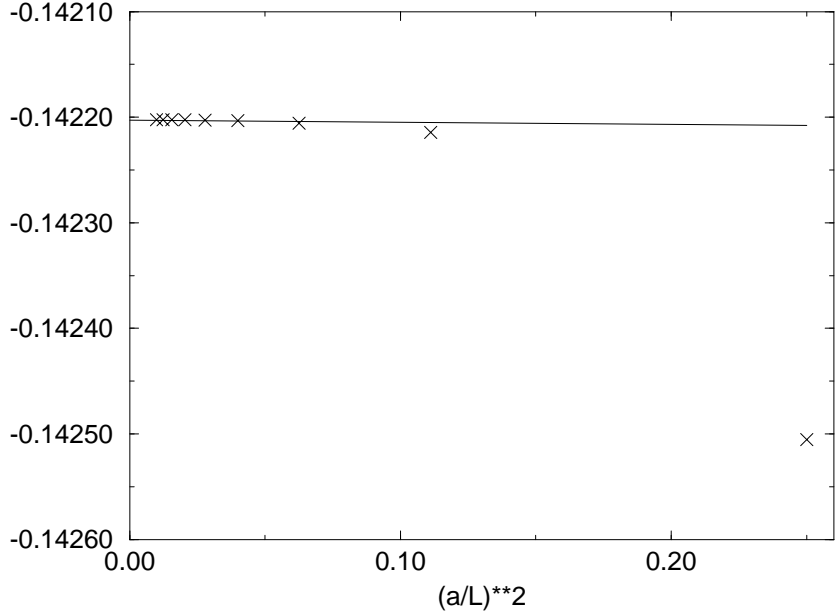}
\end{center}
\caption[]{The value of $R = R_1 + 2 R_2$ of eq.~(\ref{eq:Ri})
vs. $(a/L)^2$ for the FP action for $L/a = 2, 3, \ldots, 10$. The fit is 
$ -0.1422022 - 0.0000176\, (a/L)^2$. Note that the exact limiting value is
 $-0.1422024$.}
\label{fig:perfect}
\end{figure}

The formal RG arguments which lead to eq.~(\ref{eq:RG}) are valid in an 
infinite system. In a box whose size is comparable to the range of 
the interaction cut--off effects are generated which should go however 
to zero exponentially as the size of the system is increased. Similar
cut--off effects were observed in the correlation function of FP
operators \cite{HN-I}. The fit in fig.~1 shows this additional cut--off 
effect at $L=2$. This is a real effect which decays
rapidly and becomes part of the numerical error for $L > 3$ and it is 
related to the finite  extension of the FP action.

\end{document}